\documentclass[12pt]{article}\usepackage{latexsym,amsmath,amssymb,theorem,epsfig}
\DeclareFontFamily{OT1}{rsfs10}{}
\DeclareFontShape{OT1}{rsfs10}{m}{n}{ <-> rsfs10 }{}
\DeclareMathAlphabet{\mathscript}{OT1}{rsfs10}{m}{n}

\numberwithin{equation}{section}

\newcommand{\pt}{\partial}

\def\a{\alpha}
\def\b{\beta}
\def\g{\gamma}

\def\d{\delta}
\def\e{\epsilon}

\def\q{\theta}

\def\s{\sigma}
\def\t{\tau}

\def\G{\Gamma}

\def\gsim{ \lower .75ex \hbox{$\sim$} \llap{\raise .27ex \hbox{$>$}} }
\def\lsim{ \lower .75ex \hbox{$\sim$} \llap{\raise .27ex \hbox{$<$}} }
\def\be{\begin{equation}}
\def\ee{\end{equation}}
\def\bea{\begin{eqnarray}}
\def\eea{\end{eqnarray}}

\def \td {\tilde}
\def \rx {{\rm x}}

\def \del{\partial}
\def \a {\alpha}

\def \ci {\cite}

\def \foot {\footnote}

\def\la{\label}\def\foot{\footnote}\newcommand{\rf}[1]{\eqref{#1}}

\def \no {\nonumber}

\def \a {\alpha }

\def\foot{\footnote}

\def \ci  {\cite}
 
\theoremstyle{plain}

{\theorembodyfont{\rmfamily} }

\def \ed {\end{document}}

\begin{document}

\begin{titlepage}

\vspace{-5cm}


\vspace{-5cm}

\title{
   {\LARGE   One-loop  correction to  the energy\\ of a wavy line  string  in $AdS_5$ 
}
\\[1em] }
\author{
   E. I. Buchbinder
     \\[0.5em]
   {\it \small School of Physics M013, The University of Western Australia,}\\
       {\it \small 35 Stirling Highway, Crawley W.A. 6009, Australia}\\[0.7em]
A. A. Tseytlin\footnote{Also at Lebedev Institute, Moscow}
\\[0.5em]
   {\it \small The Blackett Laboratory, Imperial College, London SW7 2AZ, U.K.}
   }

\date{}

\maketitle

\begin{abstract}

We  consider  a computation of  one-loop $AdS_5 \times S^5$ superstring 
 correction  to the energy radiated  by the  end-point of   a  string 
 which  moves along a wavy line at the boundary of $AdS_5$ with a small transverse acceleration
(the corresponding classical solution was  described by  Mikhailov  in hep-th/0305196). 
We  also  compute  the one-loop effective action for an arbitrary small transverse string fluctuation background.  
  It is related by an analytic  continuation  to  the  Euclidean effective action
describing  one-loop correction to the expectation value of a wavy Wilson line.
We show that both the one-loop contribution to the energy and to the Wilson line 
 are controlled by the  subleading term in the strong-coupling expansion of the 
  function $B(\lambda)$ as  suggested   by Correa,  Henn,  Maldacena and  Sever in   arXiv:1202.4455.

\end{abstract}

\thispagestyle{empty}

\end{titlepage}

\section{Introduction}


In \cite{Correa:2012at} it was  suggested that there is universality in a  class of
near-BPS problems: expectation values  of small-cusp  Euclidean Wilson loop, Euclidean wavy Wilson line,
energy radiated  by a moving particle in Minkowski space.
They all are controlled by the same function $B(\lambda)$
\bea
&&B(\lambda) = \frac{\sqrt{\lambda}}{4 \pi^2} \frac{I_2 (\sqrt{\lambda})}{I_1 (\sqrt{\lambda})}
=\frac{\sqrt{\lambda}}{4 \pi^2} -\frac{3}{8 \pi^2} + \frac{3}{32 \pi^2 \sqrt{\lambda}} +
\dots \,.
\label{0.1}
\eea
In the  generalized cusp case
 the corresponding one-loop strong-coupling corrections were  already discussed in \ci{df}. 
In this paper we attempt to  further  check  this prediction  at strong coupling in the case of radiation  
and wavy line, reproducing the first subleading coefficient in \rf{0.1}  by a  direct one-loop superstring computation.
 
We shall   start in  section 2 with a 
 review of  a classical  string solution in  Minkowski-signature $AdS_5$ ending on
  a nearly-straight time-like line at the boundary. It is   described  in the static   gauge 
by  the  Mikhailov's solution \cite{Mikhailov:2003er} (see also     \cite{Kruczenski:2012aw}).
We compute the classical  string action and  the energy on this solution. 
In the remaining sections we shall  compute the 
one-loop corrections to these classical expressions.

 In section 3 we   shall  first consider the contribution from the
bosonic string fluctuations using static gauge.  Similar  computations  (see  \ci{Drukker:2000ep,kruz,mak})  
involve  several subtle points 
that we will  not directly address here (assuming cancelation of UV divergences  and  conformal anomaly
should happen   in the full theory as it does in the conformal gauge). 
In particular,  we will thus  ignore the  one-loop path integral over the $S^5$  string  modes as it should 
 contribute only to the divergences and the conformal anomaly.  
 
In section 4 we  will compute  the one-loop fermionic contributions. 
We avoid  detailed   study of   cancellation of  UV  divergences 
 by using a  heuristic (but natural) argument  of how to extract   the relevant  finite contribution. 
 Summing up  the bosonic
and the  fermionic contributions we find that the full one-loop correction  to the radiated energy 
  is indeed   proportional  to  the
next-to-leading coefficient in the expansion in  \eqref{0.1}.

Finally, in  section 5 we  will consider the Euclidean  wavy Wilson line
solution \cite{Polyakov:2000ti, Polyakov:2000jg, Semenoff:2004qr}.
The one-loop correction to the expectation value
of the wavy Wilson line  is  given by  the Euclidean one-loop effective action which
can be obtained by an analytic continuation 
 from the  corresponding Minkowski-signature  expression   computed in Sections 3 and 4. 
 As a result,  we  find  that this  one-loop correction   is also governed
by the next-to-leading coefficient in   \eqref{0.1}.

\def \rx {{x}}


\section{Classical  solution for a string ending on a wavy line}


We  shall  parametrize the worldsheet by $(\tau, \sigma)$ and $AdS_5$ by the Poincare
coordinates
\bea
&&
(X^{\mu}, Z)= (X^0, X^i, Z)\,, \qquad\qquad  \mu =0, \dots, 3\,, \quad i=1, 2, 3\,,
\nonumber \\
&&
ds^2= \frac{-(d X^0)^2 + (d X^i)^2 + dZ^2}{Z^2}\,.
\label{1.1}
\eea
%
Except for the last section, in this paper
both the worldsheet and $AdS_5$ are assumed to  have Minkowski  signature 
 with $\tau$ and $X^0$ being timelike.
We will  consider  the static gauge
\be
X^0 =\tau\,, \qquad Z=\s \,.
\label{1.2}
\ee
The Nambu-Goto action is then given by
\be
S_{cl}=-\frac{\sqrt{\lambda}}{2 \pi} \int d \t d \s \sqrt{-g}\,,
\label{1.3}
\ee
where the induced metric is 
\be
g_{\a \b} = \frac{1}{Z^2} \eta_{\mu \nu} \pt_{\a}X^{\mu}  \pt_{\b}X^{\nu}=
g_{0\a \b} + \frac{1}{\s^2} \pt_{\a}X^{i}  \pt_{\b}X^{i}\,, \quad \a, \b=1, 2
\label{1.4}
\ee
with $g_{0 \a \b}$ being the $AdS_2$ metric $g_{0 \a \b}= \frac{1}{\s^2} {\rm diag} (-1, 1 )$,

The  simplest solution  is when a string end   moves  along a  time-like straight  line at the boundary
$
X^i(\tau, \s=0)=0 .
$
We shall consider more general solution  where a string end-point moves along small deviation of that straight line. 
To leading order in
$X^i$ the  string action \rf{1.3} is given by
\be
S_{cl}=\frac{\sqrt{\lambda}}{4 \pi} \int d \t d\s \frac{1}{\s^2}[ (\pt_{\t} X^i)^2 -(\pt_{\s} X^i)^2]
\label{1.5}
\ee
and  the equation of motion  is 
\be
-\pt^2_{\t} X^i +\pt^2_{\s} X^i -\frac{2}{\s}\pt_{\s} X^i =0\,.
\label{1.6}
\ee
Its  solution  with  boundary conditions
 \be
X^i (\tau, \s=0) \equiv {\rm x}^i (\tau)= {\rm x}^i_+ (\tau) + {\rm x}^i_- (\tau)
\label{1.8}
\ee
is given by 
 \cite{Mikhailov:2003er}
\be
X^i= \rx^i_+ (\t+ \s) -\s \dot{\rx}^i_+ (\t+ \s) +
\rx^i_- (\t- \s) +\s \dot{\rx}^i_- (\t- \s) \,,
\label{1.7}
\ee
where $\rx^i_+$, $\rx^i_-$ are arbitrary 
 functions and  by dot
we denote the derivative over  $\tau$.
The solution is
uniquely defined by the boundary curve \rf{1.8} 
 the third normal derivative $(\pt_{\s}^3 X^i)\big|_{\s=0}$ \cite{Mikhailov:2003er}.
In this paper, for simplicity, we will set 
\be \label{1.77}
{\rm x}^i_+(\tau) =0 \ , \ \ \ \ \ \ \ \ \ \ \ 
{\rm x}^i_- (\tau)= {\rm x}^i (\tau) \ . \ee
We shall reserve the notation $x_i$ for $x^i_- (\tau-\s)$
using ${\rm x_i}$ for the function $ {\rm x}^i_- $ of $\tau$ only.

Let us evaluate the action and the energy on this solution in terms of the boundary
data ${\rm x}^i (\tau)$ \cite{Mikhailov:2003er, Kruczenski:2012aw}.
Using equations of motion and ignoring the
total $\tau$-derivative the classical action can be written as
\be
S_{cl}= \frac{\sqrt{\lambda}}{4 \pi} \int \frac{d \t}{\s^2}
(X^i \pt_{\s} X^i)\Big|_{\s=\e}^{\infty} =
-\frac{\sqrt{\lambda}}{4 \pi} \int \frac{d \t}{\s^2} X^i \pt_{\s} X^i\Big|_{\s=\e}\,,
\label{1.9}
\ee
where we introduced the cut-off near the boundary and assumed that $X^i$ vanishes at infinity.
Using the solution~\eqref{1.7},\rf{1.77} 
we obtain
\be
\frac{1}{\s^2} X^i \pt_{\s} X^i =-\frac{1}{\s} \rx^i (\tau-\s) \ddot{\rx}^i (\t-\s) -
\dot{\rx}^i (\t-\s) \ddot{\rx}^{i}(\t-\s)\,.
\label{1.10}
\ee
Expanding $\rx^i (\t-\s)$ near $\s=\e\to 0$ we get
\be
\frac{1}{\s^2} X^i \pt_{\s} X^i=-\frac{1}{\e} {\rm x}^i (\t) \ddot{{\rm x}}^i (\t) +
{\rm x}^i (\t) \dddot{{\rm x}}^{i} (\t)\,.
\label{1.11}
\ee
Substituting in~\eqref{1.9} and integrating by parts gives
\be
S_{cl}= -\frac{\sqrt{\lambda}}{4 \pi} \frac{1}{\e} \int d\t \ (v^i)^2 +
\frac{\sqrt{\lambda}}{4 \pi} \int d\t \ v^i a^i \,,
\label{1.12}
\ee
where
\be 
v^i = \dot{{\rm x}}^i (\t), \ \ \ \ \ \ \ \ \ \ \ \ \  a^i= \ddot{{\rm x}}^i (\t) \ee
 are the velocity and
the acceleration of the  string's end-point.
 Hence, the finite part of the action is given by
\be
S_{cl, fin} =\frac{\sqrt{\lambda}}{4 \pi} \int d\t \ v^i a^i \,,
\label{1.13}
\ee
Now let us evaluate the classical energy. We define the target space energy as
\be
E_{cl}= -\int d \s \frac{\pt L}{\pt \pt_{\t}X^0}\,,
\label{1.14}
\ee
where $L$ is the string  Lagrangian in \rf{1.3}. 
Eq. \eqref{1.14}  may be interpreted as  the energy radiated by  the end-point particle
moving with acceleration.
 To compute~\eqref{1.14}
in static gauge~\eqref{1.2} we have to restore the dependence on $X^0$
in the induced metric~\eqref{1.4}
and set $X^0=\t$ after we take the derivative. Alternatively,
since in the static gauge  we have $X^0=\t$,  the energy 
\eqref{1.14} coincides with the two-dimensional
Hamiltonian i.e.  
%
\be
E_{cl} (\tau) = \frac{\sqrt{\lambda}}{4 \pi} \int \frac{d \s}{\s^2}
[(\pt_{\t} X^i)^2 + (\pt_{\s} X^i)^2]\,.
\label{1.15}
\ee
To find $E_{cl}$ in terms of the boundary data we first differentiate~\eqref{1.15}
with respect to $\tau$,
\be
\pt_{\tau}E_{cl} (\tau)=\frac{\sqrt{\lambda}}{2 \pi} \int \frac{d \s}{\s^2}
[\pt_{\t} X^i \pt_{\t}^2 X^i + \pt_{\s} X^i \pt_{\t}\pt_{\s} X^i]\,,
\label{1.16}
\ee
 then integrate by parts and  then   use equations of motion~\eqref{1.6}. 
 As a result,  
%
\be
\pt_{\tau}E_{cl} (\tau)=
-\frac{\sqrt{\lambda}}{2 \pi}\frac{1}{\s^2} \pt_{\t}X^i \pt_{\s} X^i \Big|_{\s=\e}\,.
\label{1.17}
\ee
From the solution~\eqref{1.7} we find that
\be
-\frac{1}{\s^2} \pt_{\t}X^i \pt_{\s} X^i = \frac{1}{\s} \dot{x}^i \ddot{x}^i+(\ddot{x}^{i})^{2}\,.
\label{1.18}
\ee
Expanding near the boundary $\s=\e \to 0$ we obtain
\be
-\frac{1}{\s^2} \pt_{\t}X^i \pt_{\s} X^i|_{\s=\e}= \frac{1}{\e} \dot{{\rm x}}^i \ddot{{\rm x}}^i
- \dot{{\rm x}}^i \dddot{{\rm x}}^i =\frac{1}{\e} v^i a^i -\pt_{\t}( v^i a^i) + (a^i)^2\,.
\label{1.19}
\ee
Ignoring the divergence and the total derivative we find that the energy radiated
over some period $\Sigma$ is given by
\be
E_{cl}=\frac{\sqrt{\lambda}}{2 \pi} \int_0^{\Sigma} d\t \ (a^i)^2 =2 \pi B_0(\lambda)
\int_0^{\Sigma} d\t \ (a^i)^2\,,
\label{1.20}
\ee
where to leading order   in  large tension expansion  the coefficient $B(\lambda)$ is thus 
given by
\be
B_0(\lambda) =\frac{\sqrt{\lambda}}{4 \pi^2} \,.
\label{1.21}
\ee
The above classical consideration can be extended to the  general case
of non-linear  dependence on  $X^i$ \cite{Mikhailov:2003er, Chernicoff:2009xc, Chernicoff:2009ff}
but it will not be  discussed here. 

\section{The one-loop correction in the bosonic sector}


Let us now  consider  the one-loop corrections to  \eqref{1.12}, \eqref{1.20}.
We will
compute the one-loop  effective   action  for an arbitrary transverse background $X^i (\t, \s)$
that solves the  linearized
equations of motion~\eqref{1.6}.  Let us reserve the notation $X^{\mu}$   for the classical  background and 
denote the full  $AdS_5$ string fields  as $(Y^0, Y^i, Y^4)$, i.e. 
\be
ds^2=\frac{-(dY^0)^2 + (dY^i)^2 + (d Y^4)^2}{(Y^4)^2}\,,
\label{2.1}
\ee
where  $Y^4$ will be  the radial direction. The  $S^5$ string coordinates  will be denoted
as  $\chi^a$, $a=1, \dots, 5$. We will impose the quantum
 static gauge\foot{Like in a   light-cone  gauge, there  is no non-trivial  ghost determinant in a  static gauge: 
 this gauge  is fixed directly  in terms of string coordinates, but 
 their variation  under reparametrizations  involve  the gauge parameter  only algebraically  ($\delta X = \xi^a \del_a X$). 
 The resulting ghost determinant is ultralocal, i.e. contributes only $\delta(0)$ terms. The same remark will apply  to
 the  $\kappa$-symmetry gauge we will use. 
 }
\be
Y^0= X^0=\tau\,, \qquad \ \ \ Y^4= Z=\s \,, 
\label{2.2}
\ee
i.e. the fields $Y^0, \ Y^4$  will  not fluctuate. We will also split $Y^i$ as
\be
Y^i =X^i +  \frac{y^i}{\lambda^{1/4}} \,,
\label{2.3}
\ee
where $y^i$ are the quantum fluctuations. 
Since the classical solution  is non-zero only 
in $AdS_5$, the $S^5$   fields $\chi^a$   have only  fluctuating part  (which we also rescale 
by  $\lambda^{-1/4}$). To compute the
effective action to  one-loop order we need to expand the Nambu-Goto action
\be
S_b= -\frac{\sqrt{\lambda}}{2 \pi} \int d \t d \s \sqrt{-G} \ , \ \ \ \ \ \ \ \ \ \ \ \ \ 
G_{\a \b}=g_{\a \b}(Y) +\frac{1}{\sqrt{\lambda}}\pt_{\a} \chi^a \pt_{\b} \chi^a \,,
\label{2.5}
\ee
to quadratic order in the quantum fields $y^i$, $\chi^a$  (here 
%
 $g_{\a \b}$ is the induced metric depending on the $AdS_5$ fields $Y^m$). 

Expanding~\eqref{2.5} in powers of $y_i$ and $\chi_a$ 
we will get  the classical action, then linear terms 
which vanish since the background satisfies  the equations
of motion, and, finally, the quadratic terms on which we will concentrate.


\subsection{The contribution from $S^5$}


The  quadratic term in $\chi_a$ is 
\be
S_{S^5} =-\frac{1}{4 \pi} \int d \t d \s
\sqrt{-g(X)} g^{\a \b}(X) \pt_{\a} \chi^a \pt_\b \chi^a\,,
\label{2.7}
\ee
where $g_{\a \b} (X)$ is given by~\eqref{1.4}.

In general, we can find a coordinate system where $g_{\a \b}(X)$ is conformally flat. Then
\be
\sqrt{-g(X)} g^{\a \b}(X)=\eta^{\a \b}
\label{2.7.1}
\ee
and~\eqref{2.7} is independent of $X^i$. To perform such  change of variables in the path integral one has to use
a regularization covariant with respect to the induced metric $g_{\a \b}(X)$.
 The  integration
over $\chi^a$  will produce quadratic and logarithmic divergences 
as well as a   contribution to  conformal anomaly
(see,   e.g., \cite{Birrell}). 
On the other hand, in  critical string theory
the divergences and the conformal
anomaly are expected to cancel in the  total   expression for the partition function. 
Hence we  may ignore a   non-trivial 1-loop contribution from~\eqref{2.7}
as it should cancel when we add to it similar contributions from $AdS_5$  
modes and the fermions.\foot{Note that this cancellation is non-perturbative in $X^i$  as  it is based on a
regularization covariant with respect to
the full metric $g_{\a \b}(X)$ which cannot be imposed when we expand $g_{\a \b} (X)$ in powers of $X$.}
%
%
If we vary~\eqref{2.7} with respect to $\pt \pt_\t X^0$
to find the energy we obtain
\bea
&&\frac{\pt L_{S^5}}{\pt \pt_{\t} X^0}=-\frac{1}{4 \pi}
 \sqrt{-g(X)} \frac{\pt  g^{\a \b} (X)}{\pt \pt_{\t}X^0}T_{\a \b}(\chi)\,,
\label{2.8.1}
\\ &&
T_{\a \b} (\chi) =\pt_{\a} \chi^a \pt_{\b} \chi^a -\frac{1}{2}
g_{\a \b}(X) g^{\g \d}(X)\pt_{\g} \chi^a \pt_{\d} \chi^a \ . 
\label{2.9}
\eea
%
 Hence the above cancellation  assumption  can also be formulated as
$\langle T_{\a \b}\rangle=0$ for the  full  stress-energy tensor   of all the bosonic and fermionic contributions.
Note that on physical grounds one should not   of course expect the $S^5$  fluctuations to 
contribute  non-trivially  to the energy
since the string propagates only in $AdS_5$.\foot{This is  obvious in conformal gauge but should 
be true  also in  static gauge  assuming the conformal anomaly cancellation takes place.}

To finish this subsection let us find the propagator of the fields $\chi^a$ as we will need
it in the next section. 
Using~\eqref{2.7.1} we find that that $\chi^a$ have the action of massless fields in $AdS_2$
\be 
S_{S^5}  = \frac{1}{4 \pi} \int d \t d \s [(\pt_{\t} \chi^a)^2 - (\pt_{\s} \chi^a)^2]\,. 
\label{2.9.1}
\ee
Hence, the  propagator 
$G_0 (\t_1, \s_1; \t_2, \s_2)$ satisfies 
\be
(-\pt_{\t_1}^2 +\pt_{\s_1}^2) G_0(\t_1, \s_1; \t_2, \s_2)=\d (\t_1-\t_2) \d (\s_1-\s_2)\,.
\label{2.10}
\ee
Though this  looks like the equation in flat space, $G_0$ has to satisfy the $AdS_2$ (half-plane) 
boundary conditions
\be
G_0|_{\s_1=0}=0\,, \quad G_0|_{\s_2=0}=0\,.
\label{2.11}
\ee
The corresponding solution   is then given by
\be
G_0= -\frac{1}{2\pi} \log \frac{(\s_1- \s_2)^2 - (\t_1 -\t_2)^2}{(\s_1+ \s_2)^2 - (\t_1 -\t_2)^2} \,.
\label{2.12}
\ee
Note that this  is a function of a single variable $\ell$ which is (half)  the geodesic distance in $AdS_2$
\bea
&&G_0= -\frac{1}{2\pi} \log \frac{\ell}{\ell +1}\, , 
\label{2.12.2}\\ 
&&
\ell =\frac{1}{2} \frac{ (\s_1-\s_2)^2 - (\t_1 -\t_2)^2}{2 \s_1 \s_2} \,.
\label{2.12.1}
\eea


\subsection{The contribution from $AdS_5$}


The $AdS_5$ part  of the action   is
\be
S_{AdS_5}=-\frac{\sqrt{\lambda}}{2 \pi} \int d \t d \s \sqrt{-g (Y)}\,,
\label{2.13}
\ee
where $Y^0=X^0 =\tau$, $Y^4 =Z=\s$ and the transverse coordinates are given by \eqref{2.3}.
Then the induced metric will split as
\bea
&&g(Y)= g(X) + g(X, y)+ g (y)\,,
\label{2.14}\\
&&
g_{\a \b} (X)= g_{0 \a \b} +\frac{1}{\s^2} \pt_{\a} X^i \pt_{\b} X^i \,,
\la{2.15a} \\
&&
g_{\a \b} (X, y) \equiv h_{\a \b} =\frac{1}{\lambda^{1/4}} \frac{1}{\s^2}
[\pt_{\a} X^i \pt_{\b} y^i + \pt_{\a} y^i \pt_{\b} X^i]\,,
\label{2.15}\\
&&
g_{\a \b} (y) = \frac{1}{\sqrt{\lambda}} \frac{1}{\s^2} \pt_{\a} y^i \pt_{\b} y^i \,.
\la{2.15b}
\eea
Here $g_{0 \a \b}$ is the metric of $AdS_2$ and we used the static gauge conditions
$X^0= \tau$, $Z=\s$ on the background fields and $y^0=0, \  y^4=0$ on the quantum fields.

Now we expand~\eqref{2.13} to quadratic order in $y^i$ keeping the background $X^i$ arbitrary.
We get
\be
\sqrt{-g (Y)} =\sqrt{-g(X)} \ {\rm det}^{1/2}[ 1+ g^{-1}(X) \cdot g(y) + g^{-1}(X)\cdot h]\,,
\label{2.16}
\ee
where in the second factor we denote by $g(y)$ and $h$ the matrices
$g_{\a \b}(y)$ and $h_{\a \b}$, by $g^{-1} (X)$ the matrix inverse
to $g_{\a \b} (X)$ and by $\cdot$ the matrix multiplication. We may then   write~\eqref{2.16} as
\be
\sqrt{-g(X)} \exp\big[\frac{1}{2}{\rm tr} \log
(1+ g(X)^{-1} \cdot g(y) + g^{-1}(X) \cdot h)\big]\,,
\label{2.17}
\ee
where the trace is over the worldsheet indices $(\a, \b)$. Now we can
expand~\eqref{2.17} to quadratic order in $y$
%
\be
\sqrt{-g(Y)} = \sqrt{-g(X)}[1+ \frac{1}{2}{\rm tr} g^{-1}(X) \cdot g(y)
+\frac{1}{2}{\rm tr} g^{-1}(X) \cdot h -\frac{1}{4} {\rm tr}
(g(X) \cdot h)^2 +\frac{1}{8} ({\rm tr} g^{-1}(X) \cdot h )^2 +\dots] \,.
\label{2.18}
\ee
This is the most general expression up to terms of order $y^2$ for an arbitrary background.
The first term $\sqrt{-g(X)}$ is the classical action.
The third term (linear in $h$) is linear in $y$ and hence it  vanishes
on the equations of motion for  $X^i$. Let us consider the contribution
coming from the second term
\be
S^{(1)}_{AdS_5}=-\frac{1}{4 \pi} \int d\t d \s \frac{1}{\s^2}
\sqrt{-g(X)} g^{\a \b}(X) \pt_{\a}y^i \pt_{\b}y^i \,,
\label{2.19}
\ee
where we used $g_{\a \b}(y)$ in~\eqref{2.15a}. This term is analogous to~\eqref{2.7},
so   by   our assumption  discussed in the previous  subsection  
 its contribution should  cancel   against  other  similar contributions from the sphere and fermion terms. 
Hence, the  non-trivial   bosonic  contribution  should come  from the
last two terms in~\eqref{2.18} quadratic in the  metric $h_{\a \b}$ in \rf{2.15}.

The fourth term in~\eqref{2.18} is
\be
S^{(2)}_{AdS_5} = \frac{\sqrt{\lambda}}{8 \pi} \int d \t d \s
\sqrt{- g(X)} g^{\a \b}(X)  g^{\g \d}(X) h_{\b \g}
h_{\d \a}\,.
\label{2.20}
\ee
%
Since $h$ is linear in $X^i$, $h_{\b \g} h_{\d \a}$ is already
quadratic in $X^i$ so
we can replace $g(X)$ with the $AdS_2$ metric $g_0$.
\begin{figure}[ht]
\centering
\includegraphics[width=35mm]{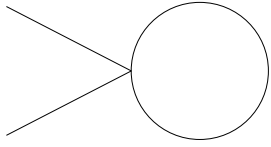}
\caption{\small The Feynman graph describing the bosonic contribution to the one-loop effective action.
The internal lines correspond to the quantum fields and the external lines are the background. Such a diagram is 
proportional to the Green's function at coinciding  points.}
\label{fig1}
\end{figure}

Similarly, the last term in~\eqref{2.18} is
\be
S^{(3)}_{AdS_5} = -\frac{\sqrt{\lambda}}{16 \pi} \int d \t d \s
\sqrt{- g(X)} g^{\a \b}(X)  g^{\g \d}(X) h_{\a \b}
h_{\g \d}\,.
\label{2.21}
\ee
Here we can also  replace $g(X)$ with $g_0$. Both~\eqref{2.20} and~\eqref{2.21}
contain only terms quadratic in $X^i$. Hence, the one-loop contribution
comes from Feynman graph   in Figure 1.

The bosonic one-loop effective action is then given by
\be
e^{i \G_b}= \int [d y^i]\ e^{i (S^{(0)}_{AdS_5}+ S^{(2)}_{AdS_5}+S^{(3)}_{AdS_5})}\,,
\label{2.22}
\ee
where $S^{(0)}_{AdS_5}$ is the free action  given by (see eq.~\eqref{2.19})
\be
S^{(0)}_{AdS_5}=\frac{1}{4 \pi}\int d \t d \s \frac{1}{\s^2}
[(\pt_{\t} y^i)^2 - (\pt_{\s} y^i)^2] \,.
\label{2.23}
\ee
We will need to find $\G_b$ to order $X^2$.
The  propagator $\langle y^i \ y^j\rangle$ comes from the free action~\eqref{2.23}
and is thus diagonal. 
Let us denote
\be
\langle y^i(\t_1, \s_1) \ y^j (\t_2, \s_2)\rangle =
2\pi G_2 (\t_1, \s_1; \t_2, \s_2) \delta^{i j}\,.
\label{2.23.1}
\ee
To find $G_2$ we change variables $y^i = \s z^i$. Then 
\be
S^{(0)}_{AdS_5}=\frac{1}{4 \pi}\int d \t d \s
\Big[(\pt_{\t} z^i)^2 - (\pt_{\s} z^i)^2 -\frac{2 (z^i)^2}{\s^2}\Big] \,.
\label{2.24}
\ee
This is the action of a massive field in $AdS_2$ with $m^2=2$. The equation for the
propagator is then 
\be
\big( -\pt_{\t_1}^2+ \pt_{\s_1}^2 -\frac{2}{\s_1^2}\big) G_2= \d (\t_1-\t_2)
\d (\s_1-\s_2) \,.
\label{2.25}
\ee
The solution to this equation satisfying the $AdS_2$ boundary conditions
$G_2|_{\s_1=0}=0, \ G_2|_{\s_2=0}=0$ is given by
\be
G_2= \frac{1}{2 \pi}\Big[-1- \frac{\s_1^2+\s_2^2 - (\t_1 -\t_2)^2}{4 \s_1 \s_2}
\log \frac{(\s_1- \s_2)^2 - (\t_1- \t_2)^2}{(\s_1+ \s_2)^2 - (\t_1- \t_2)^2}\Big]\,.
\label{2.26}
\ee
Note that it can be written as a function of the geodesic
distance parameter $\ell$  defined in ~\eqref{2.12.1}
\be
G_2=\frac{1}{2 \pi}\Big[-1 -(\ell +\frac{1}{2})\log\frac{\ell}{\ell+1}\Big]\,,
\label{2.26.1}
\ee
Now let us consider $S^{(2)}_{AdS_5}$ and $S^{(3)}_{AdS_5}$. Using the
explicit form of the matrix $h$ in~\eqref{2.15} we obtain  to quadratic
order in $X^i$ 
\bea
S^{(2)}_{AdS_5} & = & \frac{1}{8 \pi} \int d \t d \s \frac{1}{\s^2}
\eta^{\a \b} \eta^{\g \d} (\pt_{\b} X^i \pt_{\g} y^i +
\pt_{\b} y^i \pt_{\g} X^i) \
(\pt_{\d} X^j \pt_{\a} y^j +
\pt_{\d} y^j \pt_{\a} X^j)
\no \\
& = &
\frac{1}{4 \pi} \int d \t d \s \frac{1}{\s^2}
[(\pt_{\a} X^i \pt^{\a} X^j) (\pt_{\b} y^i \pt^{\b} y^j)+
(\pt_{\a} X^i \pt^{\a} y^j) (\pt_{\b} X^j \pt^{\b} y^i)]\,,
\la{227}  \\
S^{(2)}_{AdS_5} & = & -\frac{1}{4 \pi} \int d \t d \s \frac{1}{\s^2}
(\pt_{\a} X^i \pt^{\a} y^i) (\pt_{\b} X^j \pt^{\b} y^j)\,,
\label{2.27}
\eea
where  $\a,\b$ are contracted using  $\eta^{\a \b}$.
To compute $\G_b$ to order $X^2$ we  need to expand
${\rm exp}[i (S^{(2)}_{AdS} + S^{(3)}_{AdS})]$ to order $X^2$, i.e.  we need
to compute the expectation value  of the operators in~\eqref{227}, \eqref{2.27}. Since 
$\langle y^i \ y^j \rangle \sim \d^{i j}$ we can simplify~\eqref{227}, \eqref{2.27}
by taking only the diagonal contribution in $y^i y^j$. In other words, we can set
$j=i$ and sum over $i$, i.e.  keep only
\bea
(S^{(2)}_{AdS_5}+S^{(3)}_{AdS_5})_{{\rm trace}} & = &
\frac{1}{4 \pi} \int d \t d \s \frac{1}{\s^2}\sum_i
(\pt_{\a} X^i \pt^{\a} X^i)(\pt_{\b} y^i \pt^{\b} y^i)
\nonumber \\
& = &
\frac{1}{4 \pi} \int d \t d \s \frac{1}{\s^2}\sum_i
[ (\pt_{\tau} X^i)^2 - (\pt_{\s} X^i)^2] \
[ (\pt_{\tau} y^i)^2 - (\pt_{\s} y^i)^2]\,.
\label{2.28.1}
\eea
Replacing $y^i$ with $z^i=y^i/\s$ we get
\be
(S^{(2)}_{AdS_5}+S^{(3)}_{AdS_5})_{{\rm trace}}=
\frac{1}{4 \pi} \int d \t d \s \sum_i
\Big[ (\pt_{\tau} X^i)^2 - (\pt_{\s} X^i)^2\Big]
\Big[(\pt_{\tau} z^i)^2 - (\pt_{\s} z^i)^2
 -\frac{2}{\s} z^i \pt_{\s} z^i - \frac{(z^i)^2}{\s^2}\Big]
\label{2.28.2}
\ee
It is natural to define the $z$-dependent operator so that it does not
contain first derivatives. For this we integrate the third term in the second bracket
by part ignoring the terms with higher derivatives on $X$ as well as the boundary
term, thus  obtaining 
\bea
&&(S^{(2)}_{AdS_5}+S^{(3)}_{AdS_5})_{{\rm trace}}=
\frac{1}{4 \pi} \int d \t d \s \sum_i
[ (\pt_{\tau} X^i)^2 - (\pt_{\s} X^i)^2]\  {\cal O}^i\,,
\label{2.29}\\
&& \qquad \qquad 
{\cal O}^{i}=(\pt_{\tau} z^i)^2 - (\pt_{\s} z^i)^2
- \frac{2 (z^i)^2}{\s^2} \,.
\label{2.30}
\eea
The one-loop effective action is then given by
\bea
&&\G_b =
\frac{1}{4 \pi} \int d \t d \s \sum_i
[ (\pt_{\tau} X^i)^2 - (\pt_{\s} X^i)^2] \langle {\cal O}^i \rangle \,.
\label{2.31}\\
&&
\langle {\cal O}^i \rangle =\big[\pt_{\t_1} \pt_{\t_2} -\pt_{\s_1} \pt_{\s_2}
-\frac{2}{\s_1 \s_2}\big] 2\pi G_2 (\t_1, \s_1; \t_2, \s_2)\Big|_{ \t_2 \to \t_1\to \t, \ \s_2 \to \s_1\to \s}  \,,
\label{2.32}\\
\eea
where $G_2$ is given by~\eqref{2.26}. 
 Substituting $G_2$ into~\eqref{2.32}
we find
\be
\langle {\cal O}^i \rangle = C_1 + C_2 \log
\frac{(\s_1- \s_2)^2 - (\t_1- \t_2)^2}{(\s_1+ \s_2)^2 - (\t_1- \t_2)^2}\,,
\label{2.33}
\ee
where
\bea
&&
C_1=\frac{(\s_1+ \s_2)^2 (\s_1^2 +\s_2^2)
- 2 (2 \s_1^2 + 5 \s_1 \s_2 +2 \s_2^2) (\t_1-\t_2)^2 + 3 (\t_1-\t_2)^4}
{\s_1 \s_2 [(\s_1^2 +\s_2^2)^2 - (\t_1-\t_2)^2]^2}\Big|_{ \t_2 \to \t_1\to \t, \ \s_2 \to \s_1\to \s} \,,
\nonumber \\
&&
C_2=\frac{(\s_1-\s_2)^2 - 3 (\t_1-\t_2)^2}{4 \s_1^2 \s_2^2} \Big|_{ \t_2 \to \t_1\to \t, \ \s_2 \to \s_1\to \s} \,.
\label{2.34}
\eea
Taking the limit we get 
%
\be
C_1= \frac{1}{2 \s^2}\,, \qquad \qquad C_2=0\,, 
\label{2.35}
\ee
so that (for any $i$) 
\be
\langle {\cal O}^i \rangle =\frac{1}{2 \s^2}\ , 
\label{2.36}
\ee
and the bosonic one-loop effective action \eqref{2.31} is thus 
\be
\G_b =\frac{1}{8 \pi} \int d \t d \s\frac{1}{\s^2}
[ (\pt_{\tau} X^i)^2 - (\pt_{\s} X^i)^2] \,.
\label{2.37}
\ee
Note that it has  exactly the same form as 
 the classical action~\eqref{1.5}
to order $X^2$. Also note that~\eqref{2.37} is valid
for an arbitrary classical background. 
In particular, on the
Mikhailov's solution \rf{1.7} it is given by
\be
\G_b =\frac{1}{8 \pi} \int d \t \ v^i a^i\,.
\label{2.38}
\ee

Now  let us  find the one-loop correction to the energy. We can do it by
computing
\be
E_b= -\int d \s \langle
\frac{\pt L_b}{\pt \pt_{\t} X^0} \rangle \,.
\label{2.39}
\ee
For this we consider~\eqref{2.20}, \eqref{2.21} and restore
the dependence on $X^0$ in the induced metric $g(X)$. That is, we
replace in~\eqref{2.20}, \eqref{2.21} the metric $g_0$ with the
metric $\tilde{g}$ (see eq.~\eqref{1.4})
\be
\tilde{g}_{\t \t}=-\frac{1}{\s^2} (\pt_{\t} X^0)^2\,, \quad
\tilde{g}_{\t \s}=0 \,, \quad
\tilde{g}_{\s \s}=\frac{1}{\s^2}
\label{2.40}
\ee
and set $\pt_{\t} X^0=1$ after we take the derivative. Since
all the dependence on $\pt_{\t} X^0$ comes from the induced metric we get
\bea &&
\frac{\pt S^{(2)}_{AdS_5} }{\pt \pt_{\t} X^0}=\frac{\sqrt{\lambda}}{8 \pi}
\int d \t d \s \sqrt{-\td{g}} \frac{\pt \td{g}^{\mu \nu}}{\pt \pt_{\t} X^0}
[-\frac{1}{2} \td{g}_{\mu \nu} \td{g}^{\a \b} \td{g}^{\g \d}
h_{\b \g} h_{\d \a}+ 2 \td{g}^{\a \b} h_{\a \nu} h_{\b \mu}]\ , 
\label{2.41.1}
\\ &&
\frac{\pt S^{(3)}_{AdS_5}}{\pt \pt_{\t} X^0}=-\frac{\sqrt{\lambda}}{16 \pi}
\int d \t d \s \sqrt{-\td{g}} \frac{\pt \td{g}^{\mu \nu}}{\pt \pt_{\t} X^0}
[-\frac{1}{2} \td{g}_{\mu \nu} \td{g}^{\a \b} \td{g}^{\g \d}
h_{\a \b} h_{\g \d}+ 2 \td{g}^{\a \b} h_{\a \b} h_{\mu \nu}]\,.
\label{2.41.2}
\eea
Using
\be
\frac{\pt \td{g}^{\t \t}}{\pt \pt_{\t} X^0}=2 \s^2 \,, \quad
\frac{\pt \td{g}^{\t \s}}{\pt \pt_{\t} X^0}=0 \,, \quad
\frac{\pt \td{g}^{\s \s}}{\pt \pt_{\t} X^0}=0
\label{2.42}
\ee
 and substituting the explicit expression for $h$ we obtain
\bea
\frac{\pt S^{(2)}_{AdS_5}}{\pt \pt_{\t} X^0}=\frac{1}{4 \pi}
\int d \t d \s \frac{1}{\s^2} [A +2 A_{\t \t}]\,,
\qquad 
\frac{\pt S^{(3)}_{AdS_5}}{\pt \pt_{\t} X^0}=-\frac{1}{4 \pi}
\int d \t d \s \frac{1}{\s^2} [B +2 B_{\t \t}]\,,
\label{2.43}
\eea
where 
(the worldsheet indices are contracted with $\eta^{\a \b}$)
\bea
&&
A= (\pt_{\a}X^i \pt^{\a}X^j)(\pt_{\b}y^i \pt^{\b}y^j) +
(\pt_{\a}X^i \pt^{\a}y^j)(\pt_{\b}X^j \pt^{\b}y^i)\,,
\nonumber \\
&&
B= (\pt_{\a} X^i \pt^{\a} y^i)^2\,,
\nonumber \\
&&
A_{\a \b}= (\pt_{\g}X^i \pt^{\g}X^j) \pt_{\a}y^j \pt_{\b}y^i +
 (\pt_{\g}y^i \pt^{\g}y^j) \pt_{\a}X^j \pt_{\b}X^i \,,
\nonumber \\
&&
B_{\a \b} = (\pt_{\g} X^i \pt^{\g} y^i) (\pt_{\a} X^j \pt_{\b} y^j +
\pt_{\a} y^j \pt_{\b} X^j)\,.
\label{2.44}
\eea
Just like in the case of the effective action discussed above, in taking the vev in~\eqref{2.43}
only the diagonal terms in $y^i y^i$ will contribute. Thus in adding together the two expressions in 
\eqref{2.43}
we can set $j=i$ and sum over $i$. We obtain
\bea &&
(A-B)_{{\rm trace}}= \sum_i (\pt_\a X^i \pt^\a X^i) (\pt_\b y^i \pt^\b y^i)
 = \sum_i [ (\pt_{\t} X^i)^2  -(\pt_{\s} X^i)^2]\
[ (\pt_{\t} y^i)^2  -(\pt_{\s} y^i)^2]\ , \no
\\ &&
(A_{\t \t}- B_{\t \t})_{{\rm trace}} =
\sum_i (\pt_\a X^i \pt^\a X^i) \pt_\t y^i \pt_\t y^i
+\sum_i (\pt_\a y^i \pt^\a y^i) \pt_\t X^i \pt_\t X^i
\nonumber \\
&&\qquad \qquad =
- \sum_i [(\pt_{\t} X^i)^2  -(\pt_{\s} X^i)^2]\pt_\t y^i \pt_\t y^i
-\sum_i [(\pt_{\t} y^i)^2  -(\pt_{\s} y^i)^2]\pt_\t y^i \pt_\t y^i\,.
\label{2.45.2}
\eea
Note that the sum of   $ (A-B)_{{\rm trace}}$   and $    (A_{\t \t}- B_{\t \t})_{{\rm trace}} $ yields the Legendre
transform of~\eqref{2.28.1} with respect to both background and quantum fields.
Explicitly,  summing the above two equations 
we obtain the  two contributions to the energy. The first one  is 
\be
E_b= \frac{1}{4 \pi}\int d \t d \s \frac{1}{\s^2}\sum_i
[(\pt_{\t} X^i)^2 +(\pt_{\s} X^i)^2]
\langle (\pt_{\t} y^i)^2 -(\pt_{\s} y^i)^2 \rangle \,.
\label{2.46}
\ee
This term has exactly the same $y$-dependence as in~\eqref{2.28.1}.
Changing the variables $y^i=\s z^i$ and integrating by parts we obtain
\be
E_b =\frac{1}{4 \pi}\int d \t d \s \sum_i
[(\pt_{\t} X^i)^2 +(\pt_{\s} X^i)^2] \langle {\cal O}^i \rangle \,,
\label{2.47}
\ee
where the operator ${\cal O}^i$ is given  in~\eqref{2.30}. Using~\eqref{2.36} we get
\be
E_b =\frac{1}{8 \pi}\int d \t d \s \frac{1}{\s^2}
[(\pt_{\t} X^i)^2 +(\pt_{\s} X^i)^2] \,.
\label{2.48}
\ee
However, eqs.~\eqref{2.43}--\eqref{2.45.2} give rise to an additional
unexpected correction of the form
\be
\td{E}_b =\frac{1}{4 \pi}\int d \t d \s \frac{1}{\s^2}\sum_i
[(\pt_{\t} X^i)^2 -(\pt_{\s} X^i)^2]
\langle \pt_{\t} y^i \pt_{\t} y^i \rangle \,,
\label{2.49}
\ee
Let us carefully examine this term on the solution \rf{1.7}, \rf{1.77}, i.e. 
$X^i= x^i (\t - \s) +\s \dot{x}^i (\t-\s)$. First, on dimensional grounds, 
$\langle \pt_{\t} y^i \pt_{\t} y^i \rangle$ is independent of $\t$ and $\s$.
Then integrating by parts and using equations of motion we find
\be
\td{E}_b \sim \int \frac{d \s}{\s^2} \pt_{\t} (X^i \pt_{\t} X^i) -
\int d \s  \pt_{\s} \big( \frac{X^i \pt_{\s} X^i}{\s^2}\big)\,, 
\label{2.50}
\ee
where the proportionality coefficient depends on $\langle \pt_{\t} y^i \pt_{\t} y^i \rangle$.
Since the   classical  solution can be written as 
\bea
 X^i= x^i- \s x^{i \prime }\,, \ \ \ \ x^{i \prime } = \pt_{\s} x^i \ 
\label{2.51}\ , \qquad  
\pt_{\s} X^i =-\s x^{i \prime \prime}\,, \qquad
\pt_{\t} X^i =-x^{\prime i} + \s x^{i \prime \prime}\,.
\eea
we get 
\bea &&
\int d \s \pt_{\s} \big( \frac{X^i \pt_{\s} X^i}{\s^2}\big)=
\int d \s \pt_{\s} (x^{i \prime} x^{i \prime \prime} -
\frac{1}{\s} x^i x^{i \prime \prime})
\label{2.53}\\
&&
\int \frac{d \s}{\s^2} \pt_{\t}(X^i \pt_{\t} X^i)
=\int \frac{d \s}{\s^2} \pt_{\s} (x^i x^{i \prime})
-\int \frac{d \s}{\s} \pt_{\s} [(x^{i\prime})^2+ x^i  x^{i \prime \prime}]
+\int d \s \pt_{\s} (x^{i \prime} x^{i \prime \prime})
\label{2.54}
\eea
where we used that $\pt_{\t}$ acting on any function of $\t -\s$ can be
replaced with $-\pt_\s$. Now we can  integrate by parts the second term in~\eqref{2.54}.
The result will cancel the first term up to a boundary contribution. Hence, we get
\be
\int \frac{d \s}{\s^2} \pt_{\t}(X^i \pt_{\t} X^i)=-
\int d \s \pt_{\s} \big[ \frac{ (x^{i \prime})^2 +
x^i x^{i \prime \prime}}{\s} \big] +
\int d \s \pt_{\s} (x^{i \prime} x^{i \prime \prime})\,.
\label{2.55}
\ee
Substituting~\eqref{2.53} and~\eqref{2.55} in~\eqref{2.50} we obtain
\be
\td{E}_b \sim \int d \s\ \pt_{\s} \big[ \frac{ (x^{i \prime})^2}{\s}\big]
=- \frac{1}{\s}(x^{i \prime})^2\Big|_{\s =\e \to 0}\,.
\label{2.56}
\ee
Expanding near $\s =\e \to 0$  we have 
\be
x^{i \prime}=-\dot{x}^i =- \dot{{\rm x}}^i - \e \ddot{{\rm x}}^i+\dots
= -v^i - \e a^i +\dots \,,
\label{2.56.1}
\ee
so that 
\be
\td{E}_b \sim \frac{1}{\e}(v^i)^2 + 2 v^i a^i \,.
\label{2.57}
\ee
These are precisely the first two terms in~\eqref{1.19} which we have  previously 
ignored in the classical expression.
 Hence, the whole contribution $\td{E}_b$ can be ignored and the bosonic
one-loop correction is given by~\eqref{2.48}.


\section{The one-loop  contribution from the fermionic sector}


The quadratic fermionic action has the following form \cite{Drukker:2000ep, Frolov:2002av}
\be
S_f =\frac{1}{2\pi} \int d \t d\s
\big[-2 i \sqrt{-G} G^{\a \b} \bar \q \rho_\a
\nabla_\b \q +\e^{\a \b} \bar \q \rho_\a \G_* \rho_\b \theta\big]\,,
\label{3.1}
\ee
where we  imposed the   $\kappa$-symmetry  gauge 
$\theta^1 =\theta^2=\theta$ (which is ghost-free). 
In~\eqref{3.1} $\q$
is a 32-component spinor satisfying the Majorana-Weyl condition,
$G_{\a \b}$ is the induced metric which to one-loop order is simply 
the background metric $g_{\a \b} (X)$. The matrices $\rho_\a$ are 
\be
\rho_{\a}=\G_A E^A_M \pt_{\a}Y^M\,,
\label{3.2}
\ee
where $\G_A$, $A=0, \dots, 4$ are the $32 \times 32$ flat   Dirac matrices satisfying
\be
\{ \G_A, \G_B \} = 2 \eta_{A B}\,, \quad \eta_{A B}=
{\rm diag} (-1, 1, 1, 1, 1)\,.
\label{3.3}
\ee
$Y^M$ are the coordinates in $AdS_5$,   which can be replaced by their classical 
parts:  $Y^0= X^0= \t, \ Y^4=Z=\s,\ Y^i=X^i$. 
$E_M^A$ are the   $AdS_5$ vielbeins 
\be
E_M^A=\frac{1}{Z} \delta_M^A = \frac{1}{\s}\delta_M^A\,.
\label{3.4}
\ee
The matrices $\rho_\a$ are then given by
\be
\rho_\t = \frac{1}{\s} \G_0 +\frac{1}{\s} \G_i \pt_{\t}X^i\,,
\qquad\ \ \ \ \
\rho_\s = \frac{1}{\s} \G_4 +\frac{1}{\s} \G_i \pt_{\s}X^i\,.
\label{3.6.1}
\ee
They satisfy
\be
\{ \rho_{\a}, \rho_{\b}\} = 2 g_{\a \b} (X)\, .
\label{3.6.2}
\ee
For the string   $AdS_5$  background 
 the covariant derivative $\nabla_{\a}$
is given by
\be
\nabla_{\a}= \pt_{\a} +\frac{1}{4} \Omega_M^{AB} \G_A \G_B \pt_{\a} X^M\,, 
\label{3.5}
\ee
i.e. %
\bea
\nabla_{\tau}= \pt_{\tau} -\frac{1}{2 \s} \G_0 \G_4-\frac{1}{2 \s}
\G_i \G_4 \pt_{\t} X^i \,,
\qquad \qquad 
\nabla_{\sigma}= \pt_{\s}-\frac{1}{2 \s}
\G_i \G_4 \pt_{\s} X^i \,.
\label{3.6}
\eea
Finally, the matrix $\G_*$ is given by
\be
\G_* = i \G_0 \G_1 \G_2 \G_3 \G_4\,, \qquad
\G_*^2 =1\,, \qquad [\G_*, \G_A]=0\,.
\label{3.7}
\ee
We will use the basis where $\G_0$ is antisymmetric and antihermitian while
$\G_i, \ \G_4$ are symmetric and hermitian. Then $\G_*$ is antisymmetric and
hermitian.

Anticommuting Majorana spinors satisfy 
\bea
&&
\bar \q_1 \G_{A_1 A_2\dots A_n} \q_2= \bar \q_2 \G_{A_1 A_2\dots A_n} \q_1\,, \quad
n=3, 7 \,,
\nonumber \\
&&
\bar \q_1 \G_{A_1 A_2\dots A_n} \q_2=- \bar \q_2 \G_{A_1 A_2\dots A_n} \q_1\,, \quad
n\neq 3, 7 \,.
\label{3.7.1}
\eea
In particular, it follows that $\bar \q \G_{A_1 A_2 \dots A_n} \q=0$ unless
$n=3, 7$. Note that the second term in~\eqref{3.1} contains 7 gamma-matrices and,
hence, is non-zero.
\begin{figure}[ht]
\centering
\includegraphics[width=55mm]{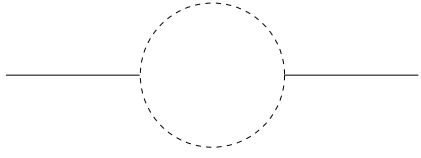}
\caption{\small The Feynman graph describing the fermionic contribution to the one-loop effective action. The internal lines
correspond to the fermionic loop and the external lines are the background. }
\label{fig2}
\end{figure}

According to our discussion in the previous section, it is natural to assume  that
the kinetic term in~\eqref{3.1} combines with~\eqref{2.7} and~\eqref{2.19} to cancel
the conformal anomaly so that  the entire non-trivial  fermionic contribution should come
from the second term in \eqref{3.1}. Expanding this  second term to order $X^2$ gives
\be
\frac{2}{\s^2} \bar \q[ i \G_0 \G_4 \G_* - \G_4 \G_i \G_* \pt_{\t}X^i +
\G_0 \G_i \G_* \pt_{\s}X^i + \G_{ij} \G_* \pt_{\t} X^i \pt_{\s}X^j]\q \,.
\label{3.8}
\ee
Note that it contains linear terms in $X^i$ and, hence to find order $X^2$  terms we have to
consider the Feynman graph   in Figure 2.
In flat space due to translational
invariance in all directions we can go to the momentum space and single
out the integral over the loop momentum. However, this cannot be done in $AdS_2$ since
there is no translational invariance in the radial direction. Therefore, we will use
a different approach.

Let is write the action \rf{3.1} as
\be
S_f =-\frac{i}{\pi} \frac{1}{2\pi} \int d \t d\s\ \sqrt{-g(X)}\
\bar \q {\cal D} \q\,,
\label{3.9}
\ee
where ${\cal D}$ is the covariant (with respect to the induced metric $g_{\a \b}(X)$)
Dirac operator
\be
{\cal D}= g^{\a \b}\rho_{\a} \nabla_{\b} +\frac{i}{2\sqrt{-g(X)}}
\e^{\a \b} \rho_{\a} \G_* \rho_{\b}\,.
\label{3.10}
\ee
%
Formally ``squaring" this operator  leads to 
\be
\hat{{\cal O}} \equiv \sqrt{-g(X)}{\cal D}^2 =
\sqrt{-g(X)}g^{\a \b}\hat \nabla_{\a} \hat \nabla_{\b} -\frac{1}{4}\sqrt{-g(X)} R^{(2)}
-\sqrt{-g(X)}\,,
\label{3.12}
\ee
where $\hat \nabla$   contains extra connection term in addition to the 2d spinor
connection.\foot{In few simple cases of string backgrounds   like 
 parallel
lines  \cite{Drukker:2000ep} or folded string \cite{Frolov:2002av}
  there is a non-trivial
rotation of fermions that puts  the  derivative $\hat \nabla$  with 
induced   connection into the form of    2d spinor covariant derivative.
That  need not be the case if the  background  is  general enough
(an example is  circular string  with 3 spins in \ci{12}). We thank S. Giombi for 
pointing out the relevance of  this issue  in the present case.}
Here $R^{(2)}$ is the curvature of the induced metric $g_{\a \b}(X)$. The first term
in~\eqref{3.12} is a  counterpart of~\eqref{2.7} and~\eqref{2.19}. By our assumption 
the contribution from this term  should cancel the similar bosonic contributions. The second
term is a total derivative and its contribution can also be ignored. 
We shall  thus assume   that,   in the first
two terms,  we can ignore the $X^i$-dependence  so that   we can effectively 
replace the metric $g (X)$ with the $AdS_2$ metric $g_0$. The non-trivial contribution 
then comes from the last term which is the square of the second term in~\eqref{3.1}.
The operator $\hat{{\cal O}}$ we will write in the form
\be
\hat{{\cal O}}= (\eta^{\a \b} \nabla_{0 \a} \nabla_{0 \b} -\frac{1}{4}
R_0 -\frac{1}{\s^2}) - (\sqrt{-g(X)} -\frac{1}{\s^2} )\equiv
\hat{{\cal O}}_0 +\hat{{\cal O}}_{int}\,.
\label{3.14}
\ee
Here $\hat{{\cal O}}_0$ is the free operator
\be
\hat{{\cal O}}_0=\eta^{\a \b} \nabla_{0 \a} \nabla_{0 \b} -\frac{1}{4}
R_0 -\frac{1}{\s^2} = -\pt_{\t}^2 +\pt_{\s}^2+ \frac{1}{\s}\G_0 \G_4 \pt_{\t}
-\frac{3}{4 \s^2}\,.
\label{3.15}
\ee
The operator $\hat{{\cal O}}_{int}$ contains interaction terms depending on the background
\be
\hat{{\cal O}}_{int}=
\sqrt{-g(X)} -\frac{1}{\s^2} =-\frac{1}{2 \s^2} [ (\pt_{\t} X^i)^2 - (\pt_{\s} X^i)^2]+\dots\,.
\label{3.16}
\ee
Hence, to order $X^2$ the operator $\hat{{\cal O}}$ can be written as
\be
\hat{{\cal O}}=\hat{{\cal O}}_0+\frac{1}{2 \s^2} [ (\pt_{\t} X^i)^2 - (\pt_{\s} X^i)^2]=
\hat{{\cal O}}_0\Big(1+ {\cal G} \cdot \frac{1}{2 \s^2} [ (\pt_{\t} X^i)^2 - (\pt_{\s} X^i)^2]\Big)\,,
\label{3.17}
\ee
where ${\cal G}$ is the Green's function of the operator $\hat{{\cal O}}_0$. The fermionic contribution to the 
one-loop
effective action is then given by\footnote{We have the coefficient 1/4 because
we integrate over the real spinors $\q$ and then square the Dirac operator.}
\bea
\G_f & = & \frac{1}{4} {\rm Tr} \log \hat{{\cal O}}=
\frac{1}{4} {\rm Tr} \log \Big(
1+ {\cal G} \cdot \frac{1}{2 \s^2} [ (\pt_{\t} X^i)^2 - (\pt_{\s} X^i)^2]\Big)
\nonumber \\
& = & \frac{1}{8} \int d \t d \s \frac{1}{\s^2}
[ (\pt_{\t} X^i)^2 - (\pt_{\s} X^i)^2]\  {\rm tr} {\cal G}(0) +\dots\,,
\label{3.18}
\eea
where ${\cal G}(0)$ is the Green's function with coincident arguments and
${\rm tr}$ is the trace over the spinor indices. 

Unfortunately, it  appears to be 
 complicated to find the Green's function of the operator~\eqref{3.15} directly. Though the operator
$\frac{1}{\sqrt{-g_0}} \hat{{\cal O}}_0$ is self-adjoint with respect to the measure
$ \int d \t d \s \sqrt{-g_0}$ its explicit form in~\eqref{3.15} is not symmetric (or hermitian)
in the usual sense. 
So we will follow an indirect approach. The equation for
${\cal G} (\t_1, \s_1; \t_2, \s_2)$ can be written as follows
\be
\Big[\rho_0^{\a} \nabla_{0 \a} +\frac{i}{\sqrt{-g_0}}\rho_{0 \tau}\rho_{0 \s} \G_*\Big]^2
{\cal G}(1;2)= \frac{1}{\sqrt{-g_0}} \d (\t_1-\t_2) \d (\s_1-\s_2)\,.
\label{3.19}
\ee
Computing the square of the first-order   operator in the left hand side of~\eqref{3.19} and multiplying it
by $ \sqrt{-g_0}=\s^{-2}$ gives precisely~\eqref{3.15}. The result of~\eqref{3.19}
can be written in the form of a convolution
\be
{\cal G}(1;2)= \int d\t_3 d \s_3 \sqrt{-g_0 (\t_3, \s_3)}\ G_f (1;3)G_f (3;2)\,,
\label{3.20}
\ee
where $G_f$ is the fermionic Green's function satisfying the first order equation
\be
\big[\rho_0^{\a} \nabla_{0 \a} +\frac{i}{\sqrt{-g_0}}\rho_{0 \tau}\rho_{0 \s} \G_*\big]G_f(1;2)
=\frac{1}{\sqrt{-g_0}} \d (\t_1-\t_2) \d (\s_1-\s_2)\,.
\label{3.21}
\ee
Using the  explicit expressions for  $\rho_{0 \a}, \nabla_{0, \a}$ we can write~\eqref{3.21}
in the form
\be
\frac{1}{\s_1}\big[-\G_0 \pt_{\t_1} + \G_4 \pt_{\s_1} +\frac{i}{\s_1}
\G_0 \G_4 \G_*\big] G_f (1;2) =\d (\t_1-\t_2) \d (\s_1-\s_2)\,.
\label{3.22}
\ee
Let us define $G_f=\sqrt{\s_1 \s_2} \ G'_f$. Then $G'_f$ satisfies
\be
[-\G_0 \pt_{\t_1} + \G_4 \pt_{\s_1} +\frac{i}{\s_1}
\G_0 \G_4 \G_*] G'_f (1;2) =\sqrt{\frac{\s_1}{\s_2}}\d (\t_1-\t_2) \d (\s_1-\s_2)\,.
\label{3.23}
\ee
Due to the $\d$-function we can replace $\sqrt{\frac{\s_1}{\s_2}}$ with 1.
Hence, we need to solve the equation
\bea
D G'_f (1;2)=\d (\t_1-\t_2) \d (\s_1-\s_2)\,
\ , \ \ \ \ \ \ \ \ \ \ \ 
D=-\G_0 \pt_{\t} + \G_4 \pt_{\s} +\frac{i}{\s}\G_0 \G_4 \G_*\,, 
\label{3.24}
\eea
where $D$ acts on the first argument. 
Let us square this  operator, i.e. 
 consider the equation
\be
D^2 G_s (1;2) =\big[-\G_0 \pt_{\t_1} + \G_4 \pt_{\s_1} +\frac{i}{\s_1}
\G_0 \G_4 \G_*\big]^2 G_s (1;2)= \d (\t_1-\t_2) \d (\s_1-\s_2)\,.
\label{3.25}
\ee
If we know the solution to this equation $G_s$ then
\be
G'_f(1;2) = D G_s (1;2)\,, \qquad G_f(1;2) = \sqrt{\s_1 \s_2} \ D G_s (1;2)\,.
\label{3.25.1}
\ee
Squaring $D$ gives explicitly 
\be
\big[-\pt_{\t_1}^2 +\pt_{\s_1}^2 -\frac{1}{\s_1^2} +\frac{1}{\s_1^2}i \G_0 \G_*\big] G_s (1;2)=
\d (\t_1-\t_2) \d (\s_1-\s_2)\,.
\label{3.26}
\ee
The matrix $i \G_0 \G_*$ is symmetric and squares to 1, so we can introduce the
orthogonal projectors
\be
{\cal P}_{\pm} =\frac{1}{2}(1\pm i \G_0\G_*) \,, \quad
{\cal P}_{\pm}^2 ={\cal P}_{\pm}\,, \quad {\cal P}_+ + {\cal P}_-=1\,, \quad
{\cal P}_+ {\cal P}_- = {\cal P}_- {\cal P}_+=0\,.
\label{3.27}
\ee
Now let us  look for a  solution in the form \be G_s = {\cal P}_+ G_+ +{\cal P}_- G_- \ . \ee
It then follows that $G_+$ and $G_-$ should satisfy the following equations
\bea
&&
(-\pt_{\t_1}^2 +\pt_{\s_1}^2 ) G_{+}(1;2)=\d (\t_1-\t_2) \d (\s_1-\s_2)\,,
\nonumber \\
&&
(-\pt_{\t_1}^2 +\pt_{\s_1}^2 -\frac{2}{\s_1^2}) G_{-}(1;2)=\d (\t_1-\t_2) \d (\s_1-\s_2)\,.
\label{3.28}
\eea
Thus  $G_+$ is the Green's function of a scalar field with $m^2=0$ which is given
in~\eqref{2.12}. Similarly, $G_-$ is the Green's function of a scalar field with
$m^2=2$ which is given in~\eqref{2.26}. Hence, we finally obtain\footnote{Since $\theta$ is a
Weyl spinor we should also introduce the Weyl projector. However, for simplicity, we
will ignore it and use the fact that in the space of Weyl spinors the trace of the
unit matrix gives 16 rather than 32.}
\be
G_f (1;2)= \frac{1}{2\pi} \sqrt{\s_1 \s_2} \ D
[{\cal P}_+ G_0 (1;2)+ {\cal P}_- G_2 (1;2)]\,, 
\label{2.23.01}
\ee
where $D$ always acts on the first argument.
Substituting $G_f$ into~\eqref{3.20} gives
\bea
&&
{\cal G}(1,2)= \frac{1}{4 \pi^2}\int d \t_3 d \s_3
\sqrt{-g_0(3)} \ \sqrt{\s_1 \s_3} \ \sqrt{\s_3 \s_2}
\nonumber \\
&&\qquad \qquad \qquad
 \times\ D [{\cal P}_+ G_0 (1;3)+ {\cal P}_- G_2 (1;3)]\
D [{\cal P}_+ G_0 (3;2)+ {\cal P}_- G_2 (3;2)] \,.
\label{3.24.01}
\eea
%
We need to evaluate ${\rm tr} {\cal G}(1;2)$ in the limit
$\t_2 \to \t_1 \to \t, \ \s_2 \to \s_1 \to \s$. 
The integral is of the same
difficulty as the loop integral in Figure 2 
 and we were not able to evaluate it explicitly.
We suggest the following indirect way to extract the finite contribution. First,  we will
write\footnote{Note that despite being a fermionic Green's function,
$G_f$ is not explicitly antisymmetric. This is due to the boundary conditions
in $AdS_2$, in particular,  to the absence of the translational invariance along
the $\s$-direction. Nevertheless, one can expect that in Feynman graphs
one can use it  as if it were  antisymmetric.}
\be
D_1 [{\cal P}_+ G_0 (1;3)+ {\cal P}_- G_2 (1;3)]=-
D_3 [{\cal P}_+ G_0 (1;3)+ {\cal P}_- G_2 (1;3)]\,.
\label{3.25.01}
\ee
where  sub-indices on $D$ indicate the arguments on which it acts. 
Then we integrate by part to form $D_3^2$ ignoring all the additional terms.
After integrating by parts we get
\be
D^2_3 [{\cal P}_+ G_0 (3;2)+ {\cal P}_- G_2 (3;2)] =2 \pi
\d (\t_1-\t_2) \d (\s_1-\s_2) \,.
\label{3.26.01}
\ee
Then 
\be
{\rm tr} {\cal G}(0)=\frac{1}{2 \pi} {\rm tr}
({\cal P}_+ G_0 + {\cal P}_- G_2)=
\frac{1}{2 \pi} \cdot 16 \cdot \frac{1}{2}(G_0+ G_2)\,.
\label{3.27.1}
\ee
From the expressions for $G_0, \ G_2$ in ~\eqref{2.12.2}, \eqref{2.26.1} we find
\be
G_0+ G_2=-1+ (\ell+1) \log \frac{\ell}{\ell +1}\,.
\label{3.28.01}
\ee
%
We see that there is a natural finite contribution -1 in~\eqref{3.28.01}.\footnote{Note that one-half of the
geodesic distance $\ell$  is a usual regulator in the point-splitting  regularization in field
theory in curved space-time,  for a review see,  e.g., ~\cite{Vassilevich:2003xt}.}
Thus the finite part of ${\rm tr} {\cal G}(0)$ is given by
\be
{\rm tr} {\cal G}(0)= -\frac{4}{\pi}\,.
\label{3.30}
\ee
Substituting this  into~\eqref{3.18} gives
\be
\G_f =-\frac{1}{2 \pi} \int d \t d \s \frac{1}{\s^2}
[(\pt_{\t} X^i)^2 - (\pt_{\s} X^i)^2]\,.
\label{3.31}
\ee
This is the final expression for the one-loop correction
to the order $X^2$ effective action  coming from  the fermionic sector.

To find the  corresponding contribution to the 
energy we will start with  the general expression for the one-loop
effective action
\be
\G_f =\frac{1}{4} {\rm Tr} \log \hat{{\cal O}} =\frac{1}{4}
{\rm Tr} \log \Big[1-{\cal G} \cdot (\sqrt{-g(X)} -\frac{1}{\s^2})\Big]\,,
\label{3.32}
\ee
and vary it with respect to $\pt_{\t} X^0$. 
Taking the functional derivative 
 gives
\be
-\frac{\pt \G_f}{\pt \pt_{\t} X^0}=\frac{1}{4}{\rm Tr}
\Big[\frac{1}{1-{\cal G}\cdot (\sqrt{-g(X)} -\frac{1}{\s^2})}
\cdot {\cal G} \cdot \frac{1}{2} \sqrt{-g(X)} g^{\t \t}(X)
\frac{\pt g_{\t \t}(X)}{\pt \pt_{\t} X^0}
\Big]\,.
\label{3.33}
\ee
Expanding~\eqref{3.33} to order $X^2$ produces two contributions. One comes
from expanding $\sqrt{-g(X)}$ in the denominator. It is proportional  to 
$(\pt_{\t}X^i)^2 -(\pt_{\s}X^i)^2$. This kind of term was discussed at the
end of the previous section in eqs.~\eqref{2.49}--\eqref{2.57} and was shown to be
irrelevant. The relevant contribution comes from expanding the numerator
$\sqrt{-g(X)} g^{\t \t}(X)$ which gives $-1- \frac{1}{2}[\pt_{\t}X^i)^2 +(\pt_{\s}X^i)^2]$.
Using
\be
\frac{\pt g_{\t \t}(X)}{\pt \pt_{\t} X^0}=-\frac{2}{\s^2}
\label{3.33.1}
\ee
we find  (ignoring the $X$-independent term)
\be
-\frac{\pt \G_f}{\pt \pt_{\t} X^0}=\frac{1}{4}{\rm Tr}
\big({\cal G} \cdot
\frac{1}{2 \s^2}[\pt_{\t}X^i)^2 +(\pt_{\s}X^i)^2]\big)\,.
\label{3.34}
\ee
The one-loop  fermionic contribution to the energy is  then given by
\be
E_f = \frac{1}{8} \int d \s \frac{1}{\s^2}
[\pt_{\t}X^i)^2 +(\pt_{\s}X^i)^2] \ {\rm tr} {\cal G}(0)
=-\frac{1}{2 \pi} \int d \s \frac{1}{\s^2}
[\pt_{\t}X^i)^2 +(\pt_{\s}X^i)^2]\,,
\label{3.36}
\ee
where we used eq.~\eqref{3.30}.

Let us now combine together  the bosonic and fermionic one-loop   contributions  
\bea
&&
\G_{1-loop}=\G_b +\G_f =-\frac{3}{8 \pi}
\int d \t d \s \frac{1}{\s^2}
[\pt_{\t}X^i)^2 -(\pt_{\s}X^i)^2] \,,
\la{3337}  \\
&&
E_{1-loop}=E_b +E_f =-\frac{3}{8 \pi}
\int  d \s \frac{1}{\s^2}
[\pt_{\t}X^i)^2 +(\pt_{\s}X^i)^2]\,.
\label{3.37}
\eea
Evaluating these expressions  on the Mikhailov's solution as 
in section 2  gives 
\bea
&&
\G_{1-loop} =-\frac{3}{8 \pi} \int d \t \ v^i a^i \,,
\la{338}  \\
&&
E_{1-loop}=-\frac{3}{4 \pi} \int d \t \ (a^i)^2 =
2\pi B_1(\lambda) \int d \t \ (a^i)^2\,,
\label{3.38}
\eea
%
where  the one-loop contribution to $B(\lambda)$ is given by (cf. \rf{1.21}) 
\be
B_1(\lambda)= -\frac{3}{8 \pi^2}\,.
\label{3.39}
\ee
This is the same as the second coefficient in~\eqref{0.1}.
We thus  provided a direct string perturbative 
check of the exact expression  ~\eqref{0.1}  proposed in 
\cite{Correa:2012at}. 


\section{One-loop correction to the expectation value of a  wavy Wilson line}


Let us now
 perform an analytic continuation along $\tau$ and $X^0$ and consider
the Euclidean worldsheet and the Euclidean $AdS_5$ space. We will also work in the 
static gauge~\eqref{1.2}.
A wavy Wilson line is a small deviation from a straight
line $X^0=\tau, \ Z=\s$ in the transverse
directions $X^i$ in Euclidean
space \cite{Polyakov:2000ti, Polyakov:2000jg, Semenoff:2004qr}.
To quadratic order in the transverse fields the Euclidean string action is
\be
S_e =\frac{\sqrt{\lambda}}{4 \pi}\int \frac{d \t d \s}{\s^2}
\big[ (\pt_{\t} X^i)^2 + (\pt_{\s} X^i)^2\big] \ , 
\label{4.1}
\ee
and the equations of motion are
\be
\pt^2_{\t} X^i +\pt^2_{\s} X^i -\frac{2}{\s}\pt_{\s} X^i =0\,.
\label{4.2}
\ee
A wavy line is a solution to these equations with the boundary condition
\be
X^i (\t, \s)\Big|_{\s=0} ={\rm x}^i (\t)\,,
\label{4.3}
\ee
where ${\rm x}^i (\t)$ is an arbitrary curve on the boundary. The precise
form of the solution is given by~\cite{Semenoff:2004qr}
\be
X^i(\t, \s) =\int \frac{d \tau' }{\pi} \ {\rm x}^i (\t') \ 
\frac{2 \s^3}{((\t-\t')^2 +\s^2)^2}\,.
\label{4.4}
\ee
Substituting it into the action gives\foot{
It is straightforward to show that~\eqref{4.5} is obtained starting with~\eqref{4.1} 
(after dropping a total derivative). 
The minus sign in front of the Euclidean action in~\cite{Semenoff:2004qr} is a misprint.}
\be
S_e= -\frac{\sqrt{\lambda}}{8 \pi^2} \int d \t d \t' \
\frac{[ (\dot{\rm x}(\t) - \dot{\rm x}(\t')]^2}{(\t -\t')^2}\,.
\label{4.5}
\ee
The leading classical term in the  expectation value of the wavy Wilson line is  then  given by
\bea
\langle W_{wavy}\rangle  = e^{-S_e} & = &
1+ \frac{\sqrt{\lambda}}{8 \pi^2} \int d \t d \t'\
\frac{[ (\dot{\rm x}(\t) - \dot{\rm x}(\t')]^2}{(\t -\t')^2} +\dots
\nonumber \\
 & = & 1 +\frac{1}{2}B_0(\lambda)
\int d \t d \t'\
\frac{[ (\dot{\rm x}(\t) - \dot{\rm x}(\t')]^2}{(\t -\t')^2} +\dots\,,
\label{4.6}
\eea
where we ignored the higher order terms in ${\rm x}^i$ and 
%
\be
B_0(\lambda)=\frac{\sqrt{\lambda}}{4 \pi^2}\,.
\label{4.7}
\ee
To the one-loop order the expectation value of $\langle W_{wavy}\rangle$ is 
\be
\langle W_{wavy}\rangle= e^{-S_e - \G_{e, 1-loop}}\,,
\label{4.8}
\ee
where $\G_{e, 1-loop}$ is the Euclidean one-loop effective action computed on the
solution~\eqref{4.4}. In the previous sections we computed the one-loop correction
to the Minkowskian effective action ~\eqref{3.37}  for an arbitrary classical background $X^i$.
Its Euclidean analog can be found by a simple analytic
continuation:
\be
\G_{e, 1-loop}=-\frac{3}{8 \pi}
\int d \t d \s \frac{1}{\s^2}
\big[\pt_{\t}X^i)^2 +(\pt_{\s}X^i)^2\big]\,.
\label{4.9}
\ee
%
Evaluating it on the solution~\eqref{4.4} gives
\be
\G_{e, 1-loop} = \frac{3}{16 \pi^2} \int d \t d \t' \
\frac{[ (\dot{\rm x}(\t) - \dot{\rm x}(\t')]^2}{(\t -\t')^2}\,.
\label{4.10}
\ee
Hence, to quadratic order in ${\rm x}^i$ we get
\be
\langle W_{wavy}\rangle = 1+ \frac{1}{2}B(\lambda)
\int d \t d \t'\
\frac{[ (\dot{\rm x}(\t) - \dot{\rm x}(\t')]^2}{(\t -\t')^2} +\dots\,,
\label{4.11}
\ee
where the one-loop correction to $B(\lambda)$ is given by
\be
B(\lambda)= B_0 + B_1 + ... \ , \ \ \ \ \ \ \   B_1=  -\frac{3}{8\pi^2} \ , 
\label{4.12}
\ee
i.e.  is the same as in~\eqref{3.39}. 
Thus we  checked   that the one-loop correction
to the expectation value of the wavy Wilson line is controlled by
the same function $B(\lambda)$ as the energy radiated by a  moving particle
in  agreement with  the claim of \cite{Correa:2012at}.


\section*{Acknowledgements}
We are  grateful  to   M. Kruczenski  for useful  discussions.
The work of E.I.B. was supported by the ARC Future Fellowship FT120100466.
The   work of A.A.T. was supported by the ERC Advanced grant No.290456
and also by the STFC grant ST/J000353/1.
E.I.B. would also like to thank Theory Group at Imperial College and Center for Theoretical Physics at 
Tomsk State Pedagogical University where the part of the work was done
for warm hospitality.


\end{document}